\documentclass[aps,prb,twocolumn,
showpacs]{revtex4}
\usepackage{ae}
\usepackage[T1]{fontenc}
\usepackage[ansinew]{inputenc}
\usepackage{amsmath}
\usepackage{amssymb}
\usepackage{graphicx}
\usepackage{color}
\usepackage[colorlinks]{hyperref}
\usepackage{epstopdf}

\begin{document}

\date{\today}

\title{Giant tunable magnetoresistance  of electrically gated graphene
ribbon with lateral interface under magnetic field.}

\author{A. M. Kadigrobov}
\affiliation{ Theoretische Physik III, Ruhr-Universit\"at Bochum,
D-44801 Bochum, Germany }

 e-mail: kadig@tp3.rub.de

  \begin{abstract}
Quantum dynamics and kinetics of  electrically gated  graphene
ribbons  with lateral n-p and e-n-p junctions under magnetic field
are investigated. It is shown that the snake-like states of
quasiparticles skipping along the n-p interface do not manifest
themselve in the main semiclassical part of the ribbon
conductance. Giant oscillations of the conductance of a ribbon
with an n-p-n junction are predicted and analytically calculated.
Depending on the number of   junctions inside the ribbon its
magnetoresistance may be controllably changed by $50\% \div 90\%$
by an extremely small change of the magnetic field or the gate
voltage.
\end{abstract}


 \maketitle

\section{Introduction}
During the last decades, great attention has been payed to
transport properties of various mesoscopic systems
\cite{Heinzel,Dittrich} such as quantum dots, quantum nanowires,
tunneling junctions and 2D electron gas based nanostructures.
Fascinating quantum mechanical phenomena arise in confined quantum
Hall systems under dc or ac currents. In particular, nonlinear
current-voltage characteristics and magnetoresistance oscillations
arise due to hopping between Landau orbits in the presence of a
random potential \cite{Tsui,Yang,Dmitriev,Zhang,Vavilov,Shi}.

Dynamics and kinetics of electrons qualitatively changes if the
quantum interference of the electron wave functions with
semiclassically large phases takes place. The most prominent and
seminal phenomenon of this type is the magnetic breakdown
 phenomenon\cite{Cohen,KaganovSlutskin,FTT} in which large
semiclassical orbits of electrons under magnetic field are coupled
by quantum tunnelling through very small areas in the momentum
space. Other systems  with analogous quantum interference are
those with multichannel reflection of electrons from sample
boundaries \cite{reflection,physica}, samples with grain
\cite{Peschanski} or twin boundaries \cite{Koshkin}. Common to all
these systems are analogous dispersion equations of electrons
which are sums of $2 \pi$ periodic trigonometric functions of
semiclassically large phases of the interfering wave functions
(see papers \cite{SKmb,Slutskin}, Section 2.3, p. 202 in
paper\cite{KaganovSlutskin}, and the rest of the above citations).
All these dispersion equations determine peculiar quasi-chaotic
spectra of the magnetic breakdown type which are gapless in the
three dimensional case.

Energy gaps in semiconductors and isolators play a crucial role in
their transport and optical properties. In modern applied physics
and device technology tunable energy gaps may be  of great
importance as they allow an effective control of operation of such
devices: transistors, photodiodes, lasers and so on.

Artificial preparation of lateral potential barriers in a two
dimensional (2D)  electron gas opens wide opportunities for
obtaining spectra with tunable energy gaps, e.g., the spectrum of
the quasiparticles skipping along an artificial barrier under
magnetic field  is a series of alternating narrow energy bands and
gaps
 the width of which $\sim \hbar \omega_H$ where $\omega_H =e H/m c$ is the
 cyclotron frequency, $m$ is the electron effective mass \cite{kang,barrier}.
These features  of the electron spectrum result in an extremely
high sensitivity of thermodynamic and transport properties of the
2D electron gas to external field: giant oscillations of the
ballistic conductance (observations of which are reported  in Ref.
\cite{kang}), nonlinear current-voltage characteristics, coherent
Bloch oscillations under a weak electric fields arise in such a
system \cite{barrier}.

Experimental discovery of two-dimensional graphene \cite{graphene}
(see also Review Papers \cite{Castro,Sarma})  has opened up fresh
opportunities for manipulation of quasiparticle dynamics and
kinetics due to peculiarities of its electronic spectrum. In
neutral one layer graphene, the Fermi energy crosses exactly the
cone points of the Fermi surface, the electron and hole dispersion
laws being
%
%
%%%%%%%%%%%%%%%%%%%%%%%%%%%%%%%%%%
\begin{eqnarray}
\varepsilon_{e,h}(p_x,p_y) = \pm v \sqrt{p_x^2+p_y^2}
\label{graphdispersion}
\end{eqnarray}
%%%%%%%%%%%%%%%%%%%%%%%%%%%%%%%%%%
%
%
Here $p_x,p_y$ are projections of the quasiparticle momentum and
$v \sim 10^{8}$ cm/s is the energy independent velocity. This
feature allows one to vary the carrier density in a wide range and
create various potential barriers by applying an external gate
voltage $V_g$. In paper \cite{Zhang1},  a widely tunable
electronic band gap was demonstrated in electrically gated bilayer
graphene.

\begin{figure}
  %%%%%%%%%%%%%%%%%%%%%%%%%%%%%%%% \centerline{\psfig{figure=zlaser2.eps,width=8cm}}
  \centerline{\includegraphics[width=8.0cm]{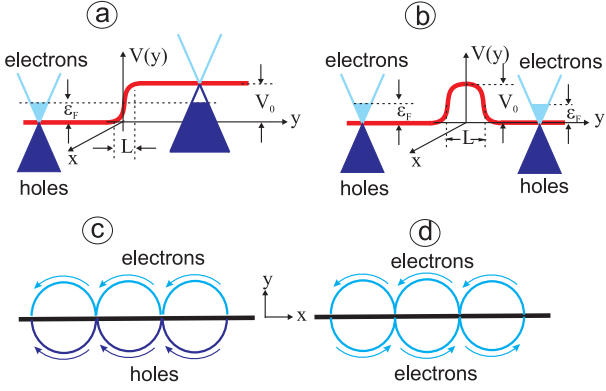}}
  %\vspace{1cm}
  \caption{Schematic presentation of n-p and n-p-n junctions. Panels a and b show
the potentials
  and the fillings the graohene bands while panels c and d show the
  classical orbits and the direction motion of electrons and holes skipping
  along the lateral junction. }
\label{lateralbarriers}
  \end{figure}

The object of this paper is to demonstrate that despite the weak
sensitivity of the quasi-particles to external electrostatic
potentials (see, e.g. Ref.\cite{Castro}), tunable bandgaps are
possible in electrically gated   graphene if one  creates lateral
barriers under magnetic field (see Fig.\ref{lateralbarriers}).
Here dynamics and kinetics of electrons skipping along
electro-hole-electron (n-p-n) and electron-hole (p-n) junctions
(see Fig.\ref{lateralbarriers}) are analytically and numerically
investigated. Giant oscillations of the conductance of a graphene
ribbon with a lateral  n-p-n junction are shown to arise
 in both the clean and dirty cases; one of the
peculiar features of the quasi-particle kinetics is   giant
magnetoresistance which takes place   every time as the Fermi
energy passes an energy gap in the electron spectrum under a
change of the magnetic field or the gate voltage.
%All quantum mechanical phenomena are enhanced in low dimensional
%systems such as a two-dimensional electron gas (2DEG), quantum
%quasi-one-dimensional wires, and so on. In particular, the
% spectrum of electrons in 2D electron gas with an artificially prepared potential
%barrier under magnetic field $H$ is a series of alternating narrow
%energy bands and gaps
%the width of which
%$\sim \hbar \omega_H$ where $\omega_H =e H/m c$ is the
%cyclotron frequency, $m$ is the electron effective mass. These
%features  of the electron spectrum results in an extremely high
%sensitivity of thermodynamic and transport properties of the 2D
%electron gas to external field: giant oscillations of the
%ballistic conductance (which where  observed in \cite{kang}),
%nonlinear current-voltage characteristics, coherent Bloch
%oscillations under a weak electric fields arise in such a system
%\cite{kang,barrier}.

%After experimental discovery of two-dimensional graphene
%\cite{graphene} a huge number of scientific papers have been
% devoted to the study of  its electronic
%properties.  \cite{Castro,Sarma}).
%
%
%%%%%%%%%%%%%%%%%%%%%%%%%%%%%%%%%%%%%%%%%%%%%%%
  %%%%%%%%%%%%%%%%%%%%%%%%%%%%%%%%%%%%%%%%%%%%%%%

\section{Dynamics of quasi-particles skipping along lateral junctions
under magnetic field.}

Let us consider  semiclassical motion of a quasiparticle moving
along
 p-n and n-p-n junctions under magnetic field as is shown in
Fig.\ref{lateralbarriers} where panels \textbf{a} and \textbf{b}
schematically present lateral electron-hole and
electron-hole-electron junctions placed along the $x$-direction;
panels \textbf{c} and \textbf{d} schematically show semiclassical
orbits of electrons skipping along the lateral junctions, the
arrow showing directions of the quasi-particle motion.

Quantum dynamics of quasi-particle (electrons and holes)in
graphene with a lateral junction is described by the 2-component
wave function $\Psi_{1,2}(x,y)$ satisfying the Schr\"{o}dinger
equation
%
%%%%%%%%%%%%%%%%%%%%%%%%%%%%%%%%%%
\begin{eqnarray}
\Big(  V(y)-\varepsilon \big)\Psi_1 +v\big(P_x+
\frac{e H}{c}y-\hbar\frac{d}{d y}\Big)\Psi_2=0;\nonumber \\
 v\Big(P_x+\frac{e
H}{c}y+\hbar\frac{d}{d y}\big)\Psi_1 +\big(  V(y)-\varepsilon
\Big)\Psi_2=0; \label{Schroedinger}
\end{eqnarray}
%%%%%%%%%%%%%%%%%%%%%%%%%%%%%%%%%%
%
where the vector potential ${\bf A}= (Hy,0,0)$ is used while
$V(y)$ is the lateral barrier potential (of the n-p or n-p-n type,
see Fig.\ref{lateralbarriers} ) extended along the $x$-direction.
Here, the axis $x$ is parallel to the sample and the barrier
junction while the $y$-axis is perpendicular to those as is shown
in Fig.\ref{lateralbarriers}; $P_x$ is the conserving projection
of generalized momentum on the lateral junction direction.

%
%%%%%%%%%%%%%%%%%%%%%%%%%%%%%%%%%%%%%%%%%%%%%%%
  \begin{figure}
  %%%%%%%%%%%%%%%%%%%%%%%%%%%%%%%% \centerline{\psfig{figure=zlaser2.eps,width=8cm}}
  \centerline{\includegraphics[width=6.0cm]{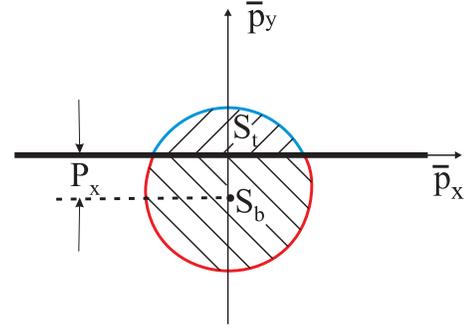}}
  %\vspace{1cm}
  \caption{Areas of the semiclassical orbits in the momentum space at
  fixed conserving momentum projection $P_x$ for quasiparticles above and below
  the junction.}
\label{Areas}
  \end{figure}
  %%%%%%%%%%%%%%%%%%%%%%%%%%%%%%%%%%%%%%%%%%%%%%%

Taking semiclassical solutions of Eq.(\ref{Schroedinger}) above
($y > 0$) and below ($y<1$) the lateral junction and matching them
at the turning points and at the junction with the use of the $2
\times 2$ scattering matrix  one finds the proper wave functions
and the quasiparticle spectrum.

I. The quasi-particle skipping along the  p-n junction (see
Fig.\ref{lateralbarriers}a,c) is in a quantum superposition of the
electron and hole edge states above ($y>0$) and below ($y<0$) the
n-p junction:
%
%
%%%%%%%%%%%%%%%%%%%%%%%%%%%%%%%%%%
\begin{eqnarray}
\hat{\Psi}_{n,P_x}(x,y)=\left(\begin{matrix} \Psi_1\\
\Psi_2
\end{matrix} \right)=e^{ix P_x/\hbar}\Big(\bar{C}_e \hat{\Psi}_{e,n,P_x}(y)
\Theta(-y)\Big)\nonumber \\
+\bar{C}_h \hat{\Psi}_{h,n,P_x}(y) \Theta(y)
 \label{superposition}
\end{eqnarray}
%%%%%%%%%%%%%%%%%%%%%%%%%%%%%%%%%%
%
%
where $n$ is the Landau number, $P_x$ is the conserving momentum
projection to the lateral junctio and $\Theta(y)$ is the unit step
function.

$\hat{\Psi}_{n,P_x}(x,y)$ are the  proper wave functions of the
Schr\"{o}dinger equation Eq.(\ref{Schroedinger}),
$\hat{\Psi}_{e}(y)$ and $\hat{\Psi}_{h}(y)$  are the semiclassical
solutions of Eq.(\ref{Schroedinger}) at $y<0$ and $y> 0$,
respectively, the both of them being normalized to the unity flux
while $|\bar{C}_{h}|^2+|\bar{C}_{e}|^2=1$.

According to  Eq.(\ref{superposition}), factors $|\bar{C}_{h}|^2$
and $|\bar{C}_{e}|^2$ are the probabilities to find the
quasiparticle above the junction (that is in the hole state) and
below it (that is in the electron state), respectively. As one
easily sees from Eq.(\ref{wavefunctions}) and
Eq.(\ref{matchequationfinal}) these factors are fast oscillating
functions of $P_x$ (on the $\hbar/R_H$ scale, $R_H$ being the
Larmour  radius of the quasiparticle cyclotron  radius).
Therefore, even a rather small change $\delta P_x \sim \hbar/R_H$
of the momenta $P_x$ greatly changes these probabilities and hence
such a change sufficiently re-distributes the probabilities to
find the quasiparticle above or below the junction.

\begin{figure}
  %%%%%%%%%%%%%%%%%%%%%%%%%%%%%%%% \centerline{\psfig{figure=zlaser2.eps,width=8cm}}
  \centerline{\includegraphics[width=6.0cm]{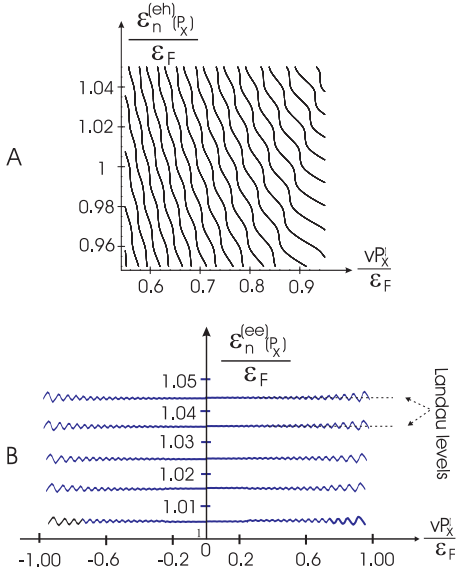}}
  %\vspace{1cm}
  \caption{The spectrum of quasiparticles skipping along the n-p interface (panel A) and
  those skipping along the n-p-n interface (panel B) Numerical calculations are performed
  for the semiclassical parameter $\eta =10^{-2}$ and the n-p-n reflection probability parameter
  $\lambda =0.2$. The spectrum of the electrons skipping along the n-p-n junction is
  an alternating series of energy gaps and bands.}
\label{ee+ehspectra}
  \end{figure}

After performing the above mentioned matching one finds the
dispersion equation (which  determines the qasiparticle spectrum
$\varepsilon_n (P_x)$) as follows:
%
%
%%%%%%%%%%%%%%%%%%%%%%%%%%%%%%%%%%
\begin{eqnarray}
D^{(eh)}\equiv \cos{\Phi_{-}^{(eh)}\left(\varepsilon, P_x\right)}
-|r^{(eh)}|\cos{\Phi_{+}^{(eh)} (\varepsilon, P_x)}=0;
\label{ehspectrum}
\end{eqnarray}
%%%%%%%%%%%%%%%%%%%%%%%%%%%%%%%%%%
%
%
where $\Phi_{+}^{(eh)} =(S_e +S_h)/2\hbar $,  $\Phi_{-}^{(eh)}
=(S_e -S_h)/2\hbar + \mu^{(eh)}$  while $S_h=2\int_{y_e}^{0}p_e dy
$ and $S_e=2\int_{y_e}^{0}p_e dy$  are the areas of the hole and
electron semiclassical orbits above and below the lateral barrier
(see Fig.\ref{Areas} in which $S_t$ and $S_b$ schematically shows
the hole and electron orbits, respectively),  $r^{(eh)}
=|r^{(eh)}|exp(i\mu^{(eh)})$  is the  reflection probability
amplitude at the junction; the turning points $y_{h,e}$ and the
integrand momenta are
%
%
%%%%%%%%%%%%%%%%%%%%%%%%%%%%%%%%%%
\begin{eqnarray}
y_h&=&\frac{c}{eH}\left(\frac{V_0 -\varepsilon}{v}-P_x\right); \nonumber \\
p_h(y)&=&\sqrt{\left(\frac{V_0-\varepsilon}{v}\right)^2-\left(P_x+
\frac{eH}{c}y\right)^2}\nonumber
\\
y_e&=&-\frac{c}{eH}\left(\frac{\varepsilon}{v}+P_x\right);  \nonumber \\
p_e(y)&=&\sqrt{\left(\frac{\varepsilon}{v}\right)^2-\left(P_x+\frac{eH}{c}y\right)^2}
\label{momentaandphi}
\end{eqnarray}
%%%%%%%%%%%%%%%%%%%%%%%%%%%%%%%%%%
%
%
At $\varepsilon \sim \varepsilon_F$ these phases
 are
%
%
%%%%%%%%%%%%%%%%%%%%%%%%%%%%%%%%%%
\begin{eqnarray}
\Phi_{\pm} \sim 1/ \eta \gg 1, \nonumber \\
\eta =\frac{ \lambda_F}{R_H }=\frac{e \hbar
H}{c}\Big(\frac{v}{\varepsilon_F}\Big)^2  \ll 1
\label{phaseestimation}
\end{eqnarray}
%%%%%%%%%%%%%%%%%%%%%%%%%%%%%%%%%%
%
%
 where
$\eta$ is the semiclassical parameter, $\lambda_F=\hbar
v/\varepsilon_F$ and $R_H =(c/eH)(\varepsilon/v)$ are the de
Broglie wave length and the Larmour radius. The numerically
calculated spectrum of quasiparticles $\varepsilon^{(eh)}_n(P_x)$
skipping along the n-p interface is present in
Fig.\ref{ee+ehspectra}, $n$ is the Landau number, $P_x$ is the
conserving momentum projection.

The reflection probability at the n-p interface may be written as
follows\cite{Castro}:
%
%
%%%%%%%%%%%%%%%%%%%%%%%%%%%%%%%%%%
\begin{eqnarray}
|r^{(eh)}\big(\varepsilon, P_x \big)|^2=
\frac{1-\sqrt{1-\big(vP_x/\varepsilon\big)^2}}{1+\sqrt{1-\big(vP_x/\varepsilon\big)^2}};
\; V_0 \gg \varepsilon_F
\end{eqnarray}
%%%%%%%%%%%%%%%%%%%%%%%%%%%%%%%%%%
%
%
where $V_0$ is the height  of the potential barrier (see
Fig.\ref{lateralbarriers})

II. The electron skipping along the n-p-n junction (see
Fig.\ref{lateralbarriers}b,d), is also in  a quantum superposition
of the electron edge states above ($y>0$) and below ($y<0$) the
junction analogous to Eq.(\ref{superposition}). However, in
contrast to the n-p junction the group velocity of the electron in
the semiclassical states above  and below the n-p-n junctions are
of the opposite signs. As a results, the electron spectrum becomes
gapped that determines peculiar properties of dynamics and
kinetics of such electrons.

In the same way as it was done for quasiparticles skipping along
the n-p junctions, matching  the electronic semiclassical wave
functions (See Appendix\ref{matching}) gives the following
dispersion equation that determines the electron spectrum
$\varepsilon_n^{(e)}(P_x)$:
%
%
%%%%%%%%%%%%%%%%%%%%%%%%%%%%%%%%%%
\begin{eqnarray}
D^{(ee)}\equiv \cos{\Phi_{+}^{(ee)}\left(\varepsilon\right)}
 -r^{(ee)}(\varepsilon,
P_x)\cos{\Phi_{-}^{(ee)}\left(\varepsilon, P_x\right)}=0;
\label{ehespectrum}
\end{eqnarray}
%%%%%%%%%%%%%%%%%%%%%%%%%%%%%%%%%%
%
%
Here $\Phi_{\pm}=c S_{\pm} /2 e \hbar  H $ while $S_{\pm} =S_1 \pm
S_2$ and $S_{1}$ and $S_{2}$ are the areas of the semiclassical
orbits
 above and below the lateral junction, respectively  (see
 Fig.\ref{lateralbarriers}). The sum and difference of the orbit areas
$S_{\pm} =S_1 \pm S_2$ are:
%
%
%%%%%%%%%%%%%%%%%%%%%%%%%%%%%%%%%%
\begin{eqnarray}
\Phi_{+}=\frac{\pi c}{2 e \hbar H} \left(\frac{\varepsilon}{v}\right)^2;\nonumber \\
\Phi_{-} = \frac{ c}{ e \hbar H}
\left(\frac{\varepsilon}{v}\right)^2 \Big\{\frac{v
P_x}{\varepsilon}\sqrt{1-\Big(\frac{v
P_x}{\varepsilon}\Big)^2}\nonumber \\ +\arcsin{\frac{\mu^{(ee)}
P_x}{\varepsilon}}\Big\}+\mu^{(ee)}
\end{eqnarray}
%%%%%%%%%%%%%%%%%%%%%%%%%%%%%%%%%%
%
%
Factor $|r^{(ee)}|^2$ is the probability of reflection at the
n-p-n junction, $\mu^{(ee)}$ is the phase of its probability
amplitude. For the sake of simplicity, one may use  the reflection
probability  in the following form\cite{Castro}:
%
%
%%%%%%%%%%%%%%%%%%%%%%%%%%%%%%%%%%
\begin{eqnarray}
r^2\left(P_x\right)=\frac{\lambda^2\left(v
P_x/\varepsilon\right)^2}{1-\left(1-\lambda^2\right)\left(v
P_x/\varepsilon\right)^2}; \;\; \lambda=\frac{V_0 L}{\hbar v}
\label{r}
\end{eqnarray}
%%%%%%%%%%%%%%%%%%%%%%%%%%%%%%%%%%
%
%
which is valid at $\lambda \ll 1$. Here $V_0$ and $L$ are the
height and the width  of the potential $V(y)$ (see
Fig.\ref{lateralbarriers}). The numerically calculated spectrum of
electrons $\varepsilon^{(ee)}_n(P_x)$  skipping along the n-p-n
interface is present in Fig.\ref{ee+ehspectra}, $n$ is the Landau
number, $P_x$ is the conserving momentum projection.

Despite dispersion equations  Eq.(\ref{ehspectrum}) and
Eq.(\ref{ehespectrum}) look much alike they determine
qualitatively different spectra: the former spectrum is gapless
(see Fig.\ref{ee+ehspectra}A) while the latter one is gapped (see
Fig.\ref{ee+ehspectra}B). As one readily sees from
Eq.(\ref{ehespectrum}) the energy gaps are determined by the
condition $$|\cos{\Phi_{+}}(\varepsilon)| \geq |r^{(ee)}|;$$ On
the other hand, one may get the necessary condition of solvability
of Eq.(\ref{ehespectrum}) $|\cos{\Phi_{-}}(\varepsilon, P_x)| <
|r^{(eh)}|$ at any energy by varying $P_x$ that provides the
gapless spectrum.

%
%%%%%%%%%%%%%%%%%%%%%%%%%%%%%%%%%%%%%%%%%%%%%%%
  %%%%%%%%%%%%%%%%%%%%%%%%%%%%%%%%%%%%%%%%%%%%%%%

In order to explicitly calculate the density of states (DOS) it is
convenient to use the approach developed by Slutskin for analogous
spectra of electrons under magnetic breakdown conditions
\cite{KaganovSlutskin}. Below,  calculations of DOS for the
gapped spectrum Eq.(\ref{ehespectrum})  are presented.

Using Eq.(\ref{ehespectrum}) and the identity
%
%
%%%%%%%%%%%%%%%%%%%%%%%%%%%%%%%%%%
\begin{eqnarray}
\sum_n\delta(\varepsilon -\varepsilon_n) \label{identity}
=\Big|\frac{\partial D^{(ee)}}{\partial \varepsilon}
\Big|\delta(D^{(ee)})
\end{eqnarray}
%%%%%%%%%%%%%%%%%%%%%%%%%%%%%%%%%%
%
%
one transforms  DOS
%
%
%%%%%%%%%%%%%%%%%%%%%%%%%%%%%%%%%%
\begin{eqnarray}
\nu(\varepsilon)=\frac{1}{2 R_H}\sum_n
\int_{-\varepsilon/v}^{\varepsilon/ v}
\delta\left[\varepsilon-\varepsilon_n(P_x)\right] \frac{d
P_x}{2\pi \hbar} \label{NuGeneral}
\end{eqnarray}
%%%%%%%%%%%%%%%%%%%%%%%%%%%%%%%%%%
%
%
into the form
%
%
%%%%%%%%%%%%%%%%%%%%%%%%%%%%%%%%%%
\begin{eqnarray}
\nu(\varepsilon)=\frac{1}{2 R_H}\sum_n
\int_{-\varepsilon/v}^{\varepsilon/ v}\Big|\frac{\partial
D^{(ee)}}{\partial \varepsilon}\Big|
\delta\left[D^{(ee)}\big(\varepsilon,P_x\big)\right] \frac{d
P_x}{2\pi \hbar} \label{NuD}
\end{eqnarray}
%%%%%%%%%%%%%%%%%%%%%%%%%%%%%%%%%%
%
%

As one sees from Eq.(\ref{ehespectrum}) the integrand here is a $2
\pi$-periodic function of $\Phi_{-}$ and hence it can be expanded
into the Fourier series as follows:
%
%
%%%%%%%%%%%%%%%%%%%%%%%%%%%%%%%%%%
\begin{eqnarray}
\nu(\varepsilon)=\frac{1}{2 R_H}\sum_{k=-\infty}^{\infty}
\int_{-\varepsilon/v}^{\varepsilon/ v}B_k\left(\varepsilon,
P_x\right)e^{ik\Phi_{-}\left(\varepsilon, P_x\right)} \frac{d
P_x}{2\pi \hbar} \label{NuFurier}
\end{eqnarray}
%%%%%%%%%%%%%%%%%%%%%%%%%%%%%%%%%%
%
%
where $B_k\left(\varepsilon, P_x\right)$ are amplitudes of the
Fourier harmonics.

As at  $\varepsilon \sim \varepsilon_F $ one has $\Phi_{-}\gg 1$
(see Eq.(\ref{phaseestimation})) the exponents in
Eq.(\ref{NuFurier}) are fast oscillating functions  while the
Fourier coefficients are smooth functions of $P_x$ (on the scale
$\hbar /R_H \ll p_F$). Therefore, the term with $k=0$ gives the
main contribution to DOS:
%
%
%%%%%%%%%%%%%%%%%%%%%%%%%%%%%%%%%%
\begin{eqnarray}
\nu(\varepsilon)=\frac{1}{2 R_H}\sum_{k=-\infty}^{\infty}
\int_{-\varepsilon/v}^{\varepsilon/ v}B_0\left(\varepsilon,
P_x\right) \frac{d P_x}{2\pi \hbar} \label{NuFurier0}
\end{eqnarray}
%%%%%%%%%%%%%%%%%%%%%%%%%%%%%%%%%%
%
%
where the Fourier factor $B_k$ at $k=0$ is
%
%
%%%%%%%%%%%%%%%%%%%%%%%%%%%%%%%%%%
\begin{eqnarray}
B_0=\int_{-\pi}^{\pi}\Big | \frac{\partial \Phi_{+}}{\partial
\varepsilon} \sin{\Phi_{+}}
-\frac{\partial \Phi_{-}}{\partial \varepsilon} \sin{\omega}\Big| \nonumber \\
\times
\delta\Big[\cos{\Phi_{+}(\varepsilon)}-|r^{(ee)}(P_x)|\cos{\omega}\Big]\frac{d
\omega}{2 \pi} \label{B0}
\end{eqnarray}
%%%%%%%%%%%%%%%%%%%%%%%%%%%%%%%%%%
%
%
%
While writing $B_0$ the explicit form of $D^{(ee)}$ (which is
given by Eq.(\ref{ehespectrum})) was used.

Carrying out integration in Eq.(\ref{B0}) and inserting the result
in Eq.(\ref{NuFurier0}) one obtains DOS as follows:
%
%
%
%%%%%%%%%%%%%%%%%%%%%%%%%%%%%%%%%%
\begin{eqnarray}
&\nu\left(\varepsilon\right)&= \frac{\big|\sin{ \Phi_{+}(\varepsilon)}\big|}{\hbar v}\nonumber \\
&\times&\int_{-\varepsilon/v}^{\varepsilon/v}\frac{\Theta\Big[|r^{(ee)}|^2(P_x)-\cos^2{
\Phi_{+}(\varepsilon)}\Big]}{\sqrt{|r^{(ee)}|^2(P_x)-\cos^2{
\Phi_{+}(\varepsilon)}}}\frac{d P_x}{2\pi \hbar} \label{NuFinal}
\end{eqnarray}
%%%%%%%%%%%%%%%%%%%%%%%%%%%%%%%%%%
%
%
Here $\Theta[...]$ is the unit step function.

\begin{figure}
  %%%%%%%%%%%%%%%%%%%%%%%%%%%%%%%% \centerline{\psfig{figure=zlaser2.eps,width=8cm}}
  \centerline{\includegraphics[width=8.0cm]{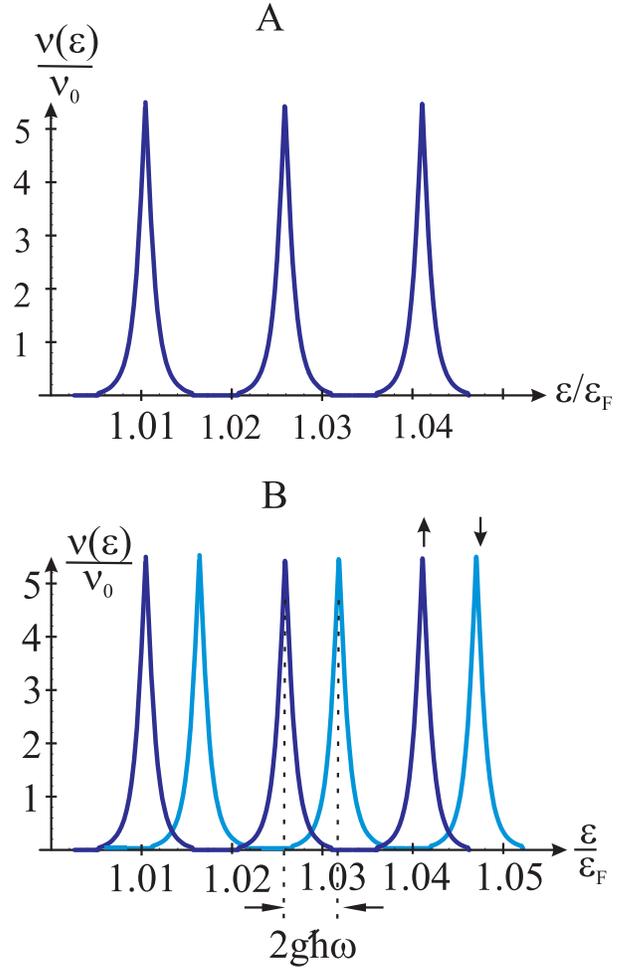}}
  %\vspace{1cm}
  \caption{Density of states for electrons skipping along the n-p-n
  interface(panel A)
  normalized to the one in the absence of magnetic field $\nu_0 =4 \pi m/(2 \pi \hbar)^2$.
Panel B shows the Zeeman split of DOS.
 Numerical calculations are performed for the semiclassical parameter $\eta =10^{-2}$
  and the n-p-n reflection probability parameter
  $\lambda =0.2$}
\label{DOSplusSpin}
  \end{figure}

The result of numerical calculations of DOS with the use of
Eq.(\ref{NuFinal}) and  Eq.(\ref{r}) for the semiclassical
parameter $\eta = 10^{-2}$ and $\lambda=0.2$ is presented in
Fig.\ref{DOSplusSpin}. As one sees  in  Fig.\ref{ee+ehspectra} and
Fig.\ref{DOSplusSpin}  the quantum interference of the edge states
above and below the n-p-n junction (see Fig.\ref{lateralbarriers})
results in arising of alternating series of energy gaps and energy
bands which produce narrow peaks in  the density of states.
Fig.\ref{DOSplusSpin}B shows the density of states caused by the
Zeeman splitting where $g$ is the gyromagnetic coefficient.
%
%
%%%%%%%%%%%%%%%%%%%%%%%%%%%%%%%%%%%%%%%%%%%%%%%
  %%%%%%%%%%%%%%%%%%%%%%%%%%%%%%%%%%%%%%%%%%%%%%%

Such a dramatic transformation of the quasi-particle spectrum has
to show itself in various prominent effects in optic and kinetic
properties. In the next section transport properties of both the
clean and dirty graphene samples are analyzed.
\section{Current along p-n junction under magnetic field. \label{currenteh}}

In this section the total current  flowing inside the stripe $- 2
R_H^{(e)} \leq y 2\leq R_H^{(h)}$ around the p-n junction is
calculated where $R_H^{(e)}=( c /e H)(\varepsilon_F/v)$ and
$R_H^{(h)}=( c /e H)((V_0-\varepsilon_F)/v)$ are the Larmour radii
of electrons (e) and holes (h);

\begin{figure}
  %%%%%%%%%%%%%%%%%%%%%%%%%%%%%%%% \centerline{\psfig{figure=zlaser2.eps,width=8cm}}
  \centerline{\includegraphics[width=8.0cm]{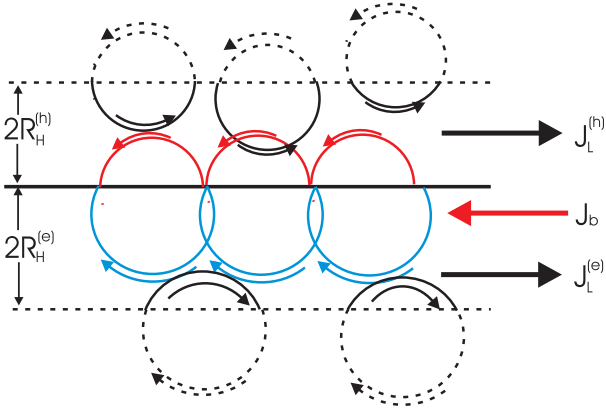}}
  %\vspace{1cm}
  \caption{Schematic presentation of currents flowing in the vicinity of the n-p
  interface. Quasiparticles which are in the quantum electron-hole superposition
 are  delocalized along the interface and  carry current $J_b$
inside  stripes of the widths  $2 R_H^{(h)}$ above the junction
and $2 R_H^{(e)}$ below the interface. Electrons and holes in the
Landau states create currents $J_L^{(e)}$ and $J_L^{(h)}$ in the
same  stripes because only parts of their closed orbits are inside
them. The later currents flow in the opposite direction to current
$J_b$ exactly compensating it in the absence of the bias voltage.}
\label{p-ncurrents}
  \end{figure}

It is easy to see that there are two types of quasiparticle states
inside this stripe: they are states of quasiparticles  which
interact with the lateral junction that delocalized them in the
junction direction, and those in which quasiparticles do not touch
the junction (the Landau states - the quasiparticles move along
closed semiclassical orbits). As only  parts of the closed orbits
are inside the stripe  these quasiparticles create finite currents
in the stripe below and above the junction. This situation is
schematically shown in Fig.\ref{p-ncurrents}.
%
%%%%%%%%%%%%%%%%%%%%%%%%%%%%%%%%%%%%%%%%%%%%%%%
  %%%%%%%%%%%%%%%%%%%%%%%%%%%%%%%%%%%%%%%%%%%%%%%
%
%

In other words, the edge states partly replace the Landau states
which would be inside the stripe in the absence of the junction
that creates an imbalance between the Landau states. As a result,
compensating currents of quasiparticles on  closed orbits arise
which flow in the opposite direction to the edge state current.
%
%%%%%%%%%%%%%%%%%%%%%%%%%%%%%%%%%%%%%%%%%%%%%%%
  \begin{figure}
  %%%%%%%%%%%%%%%%%%%%%%%%%%%%%%%% \centerline{\psfig{figure=zlaser2.eps,width=8cm}}
  \centerline{\includegraphics[width=8.0cm]{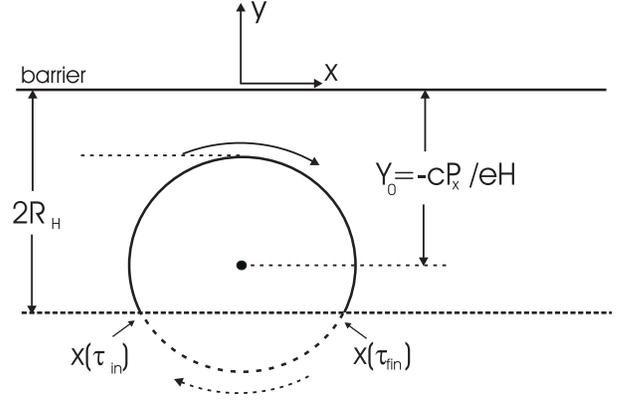}}
  %\vspace{1cm}
  \caption{A semiclassical closed orbit of an electron in the Landau state. The electron
  moving along the part of the orbit inside the stripe of the width
  $2R_H^{(e)}$} at the junction (shown by solid line) contributes to the current
  flowing inside the stripe. $x(\tau_{in})$ and $x(\tau_{fin})$
  are the initial and final  x-coordinates of this motion.
\label{eorbit}
  \end{figure}
  %%%%%%%%%%%%%%%%%%%%%%%%%%%%%%%%%%%%%%%%%%%%%%%
%
%

Let us firstly calculate the current $J_b$  carried by
quasiparticles in the edge states which flows from the right
reservoir under bias voltage $V$ to the left one under voltage
$V=0$. This current may be written as
%
%
%%%%%%%%%%%%%%%%%%%%%%%%%%%%%%%%%%
\begin{eqnarray}
J_b^{(eh)}&=&e \sum_n
\int_{-\varepsilon^{(eh)}_n/v}^{\varepsilon_n/v}
v_x\left(P_x, \varepsilon^{(eh)}_n\left(P_x\right)\right) \nonumber \\
 &\times&
f_0\Big(\varepsilon_n\left(P_x\right) +e V\Big)\frac{d P_x}{2\pi
\hbar} \label{ehbarriercurrent1}
\end{eqnarray}
%%%%%%%%%%%%%%%%%%%%%%%%%%%%%%%%%%
%
%
which may be re-written as
%
%
%%%%%%%%%%%%%%%%%%%%%%%%%%%%%%%%%%
\begin{eqnarray}
J_b^{(eh)}&=&e \int d \varepsilon f_0\left(\varepsilon +e V\right)
\int_{-\varepsilon/v}^{\varepsilon/v}\frac{d
P_x}{2\pi \hbar}\nonumber \\
 &\times&  v_x\left(\varepsilon,
P_x\right)\Big|\frac{\partial D^{(eh)}}{\partial \varepsilon}
\Big| \delta\Big[D^{(eh)}\left(\varepsilon, P_x\right) \Big]
 \label{ehbarriercurrent2}
\end{eqnarray}
%%%%%%%%%%%%%%%%%%%%%%%%%%%%%%%%%%
%
%
where $D^{(eh)}$ is defined in Eq.(\ref{ehspectrum}).

Using the same approach as in Subsection \ref{ballisticsubsection}
one finds the current carried by  quasiparticles delocalized along
the p-n junction as follows:
%
%
%%%%%%%%%%%%%%%%%%%%%%%%%%%%%%%%%%
\begin{eqnarray}
J_b^{(eh)}&=& -\frac{c}{\pi \hbar H}\int d \varepsilon
f_0\left(\varepsilon +e
V\right) \nonumber \\
&\times&\Big\{ \int_{-\varepsilon/v}^{\varepsilon/v} \frac{d
P_x}{2\pi \hbar} p_y^{(e)} + \int_{-(V_0
-\varepsilon/v)}^{(V_0-\varepsilon)/v}\frac{d P_x}{2\pi \hbar}
 p_y^{(h)} \Big \};
 \label{ehbarriercurrent3}
\end{eqnarray}
%%%%%%%%%%%%%%%%%%%%%%%%%%%%%%%%%%
%
%
Here $ p_y^{(e,h)}$ are the $y$-projections of the  electron and
hole momenta inside the electron and hole parts of the
electrically gated graphene in the absence of magnetic field:
%
%
%%%%%%%%%%%%%%%%%%%%%%%%%%%%%%%%%%
\begin{eqnarray}
 p_y^{(e)}&=&\sqrt{\Big (\frac{\epsilon}{v}\Big)^2 -P_x^{2}}\nonumber \\
 p_y^{(h)}&=&\sqrt{\Big (\frac{V_0-\epsilon}{v}\Big)^2 -P_x^{2}}
 \label{momenta}
\end{eqnarray}
%%%%%%%%%%%%%%%%%%%%%%%%%%%%%%%%%%
%
%

As it follows from Eq.(\ref{ehbarriercurrent3}) the current of
quasi-particles interacting with the junction does not depend on
its transparency and is a sum of the electron and hole edge state
currents flowing in the same direction.  These edge state currents
flows inside two  stripes:  $-2 R_H^{(e)} < y <0$  and one $0< 2
R_H^{(h)} < y <0$.

As it was said above there are two other additional currents
inside the same stripe around the junction flowing in the opposite
direction to the current carried by the edge states
Eq.(\ref{ehbarriercurrent3}).

Below,  the current  carried by electrons in the Landáu states
inside the stripe $-2 R_H^{(e)} < y <0$ is calculated.

 The current density is written as follows:
%
%
%%%%%%%%%%%%%%%%%%%%%%%%%%%%%%%%%%
\begin{eqnarray}
j_L^{(e)} (\textbf{r}_0)=e
Tr\Big\{\delta(\hat{\textbf{r}}-\textbf{r}_0)
f_0\big(\hat{{\cal H}_0}\big)\hat{v}_x \nonumber \\
+ f_0\big(\hat{{\cal H}_0}\big)\hat{v}_x
\delta(\hat{\textbf{r}}-\textbf{r}_0)\Big\}
 \label{currentdensity}
\end{eqnarray}
%%%%%%%%%%%%%%%%%%%%%%%%%%%%%%%%%%
%
where the velocity operator $\hat{v}_x $ is
%
%
%%%%%%%%%%%%%%%%%%%%%%%%%%%%%%%%%%
\begin{eqnarray}
\hat{v}_x=\frac{i}{\hbar}\left[\hat{{\cal H}}_0,\hat{x}\right]
 \label{velocityoperator}
\end{eqnarray}
%%%%%%%%%%%%%%%%%%%%%%%%%%%%%%%%%%
%
and $\hat{{\cal H}}_0$ is the Hamiltonian corresponding to the
Schr\"{o}dinger equation Eq.(\ref{Schroedinger}) in the absence of
the junction, $V(y) \equiv 0$.

Using Eq.(\ref{currentdensity}) one finds the current  inside the
stripe in the semiclassical approximation as follows:
%
%
%%%%%%%%%%%%%%%%%%%%%%%%%%%%%%%%%%
\begin{eqnarray}
J_L^{(e)} = e \int_{\varepsilon/v}^{3 \varepsilon/v}\frac{d P_x
}{2 \pi \hbar}\int \frac{dP_y}{2 \pi \hbar} \int_{-2
\widetilde{R}^{(e)}_H} ^{y_1}d y \nonumber \\
\times f_0\Big[{\cal H}_0(P_x + \frac{e H}{c}y, P_y)\Big]
v_x\Big[{\cal H}_0(P_x + \frac{e H}{c}y, P_y)\Big];
 \label{ehcurrentcloseorbits}
\end{eqnarray}
%%%%%%%%%%%%%%%%%%%%%%%%%%%%%%%%%%
%
Here the argument of the Fermi distribution function $f_0$ is the
classical Hamiltonian of the graphene under magnetic field while
${\cal H}_0 (\textbf{p})=v\sqrt{p_x^2 +p_y^2}$ is the classical
Hamiltonian of graphene at $H=0$ and $P_x, P_y$ are the
projections of the electron generalized momentum; $y_1 = - c
P_x/eH +R_H^{(e)}$ is the turning point nearest to the junction,
 $R_H^{(e)}= (c/eH)(\varepsilon/v)$  is the Larmour radius at
 fixed electron energy
$\varepsilon$; the limits of integration with respect to $P_x$ are
determined by the condition that the turning point $y_1$  is
inside the stripe, $-2\widetilde{R}^{(e)}_H  \leq y_1 \leq 0$.

It is convenient to insert  new variables $P_x, P_y \rightarrow
\varepsilon, \tau$ where $\tau$ is the time of motion along the
classical electron orbit. In this variables the equation of
electron motion  under magnetic field is the standard Hamilton
equation:
%
%%%%%%%%%%%%%%%%%%%%%%%%%%%%%%%%%%
\begin{eqnarray}
\frac{d \textbf{p}}{d \tau} =\frac{e}{c}\Big[\textbf{v}\times
\textbf{H}\Big]
 \label{Hamiltonequation}
\end{eqnarray}
%%%%%%%%%%%%%%%%%%%%%%%%%%%%%%%%%%
%
where $\textbf{p}=(P_x+ (e H/c)y, P_y)$.

Inserting the new variables in Eq.(\ref{ehcurrentcloseorbits}) and
using Eq.(\ref{Hamiltonequation}) one finds the current of the
Landau electrons inside the stripe as follows:
%
%%%%%%%%%%%%%%%%%%%%%%%%%%%%%%%%%%
\begin{eqnarray}
J_L^{(e)}=\frac{ e}{(2 \pi \hbar)^2}\frac{c}{e H}
\int_{\varepsilon/v}^{3 \varepsilon/v} dP_x \nonumber \\
 \int d
\varepsilon
f_0(\varepsilon)\left(p_y(\tau_{in})-p_y(\tau_{fin})\right)
 \label{ehcurrentclose1}
\end{eqnarray}
%%%%%%%%%%%%%%%%%%%%%%%%%%%%%%%%%%
%
where $p_y (\tau) =-(eH/c)x(\tau)$ according to
Eq.(\ref{Hamiltonequation}) and $x(\tau_{in, fin})$ are the
initial and final $x$-coordinates of motion of the electron  along
its orbit (see Fig.\ref{eorbit}). It is easy to see that
 $$p_y
(\tau_{fin})=-p_y (\tau_{in})
=\sqrt{\left(\frac{\varepsilon}{v}\right)^2-\left(P_x-2\frac{\varepsilon}{v}\right)^2};$$
Inserting this equation in Eq.(\ref{ehcurrentclose1}) one finally
finds the current of  electrons in the Landau state inside the
stripe $-2 R_H ^{(e)}\leq y \leq 0$ as follows:
%
%
%%%%%%%%%%%%%%%%%%%%%%%%%%%%%%%%%%
\begin{eqnarray}
J_L^{(e)}= \frac{c}{\pi \hbar H}\int d \varepsilon
f_0\left(\varepsilon \right) \int_{-\varepsilon/v}^{\varepsilon/v}
\frac{d P_x}{2\pi
\hbar}\sqrt{\left(\frac{\varepsilon}{v}\right)^2-P_x^2};
 \label{eLandaucurrentElectrons}
\end{eqnarray}
%%%%%%%%%%%%%%%%%%%%%%%%%%%%%%%%%%
%

Performing analogous calculations for the current $J_L^{h}$
carried by holes in Landau states inside the stripe $0 \leq y\leq
2 R_H ^{(h)}$ one gets
%
%
%%%%%%%%%%%%%%%%%%%%%%%%%%%%%%%%%%
\begin{eqnarray}
J_L^{(h)}&= &\frac{c}{\pi \hbar H} \int d \varepsilon
f_0\left(\varepsilon +e V\right) \nonumber \\
&\times& \int_{-(V_0 -\varepsilon/v)}^{(V_0-\varepsilon)/v}\frac{d
P_x}{2\pi \hbar} \sqrt{\Big (\frac{V_0-\epsilon}{v}\Big)^2
-P_x^{2}};
 \label{hLandaucurrentElectrons}
\end{eqnarray}
%%%%%%%%%%%%%%%%%%%%%%%%%%%%%%%%%%
%

Comparing Eq.(\ref{eLandaucurrentElectrons}) and
Eq.(\ref{hLandaucurrentElectrons})  with
Eq.(\ref{ehbarriercurrent3}) one sees that the
 currents of quasiparticles in the Landau states $J^{(e,h)}_L$ and the one carried by
 electrons in the edge states , $J_b^{(eh)}$, flow  in the opposite
 directions
  being modulo equal  in the absence of the bias voltage, $V=0$.

Summing the currents  given by
Eq.(\ref{ehbarriercurrent3},\ref{eLandaucurrentElectrons},\ref{hLandaucurrentElectrons})
and expanding the Fermi function with respect to $e V/k T \ll 1$
one finds the total current $J_{total}^{(eh)} = J_b^{(eh)}
+J_L^{(e)}+J_L^{(h)}$ flowing inside the stripe $- 2 R_H ^{(e)}
\leq y\leq 2 R_H ^{(h)}$ biased by the voltage drop $V$ as
follows:
%
%
%%%%%%%%%%%%%%%%%%%%%%%%%%%%%%%%%%
\begin{eqnarray}
J_{total}^{(eh)}&=& -\frac{e c}{\pi \hbar H}\int d \varepsilon
\frac{d f_0(\varepsilon)}{d \varepsilon} \Big\{
\int_{-\varepsilon/v}^{\varepsilon/v} \frac{d P_x}{2\pi \hbar}
\sqrt{\Big (\frac{\epsilon}{v}\Big)^2 -P_x^{2}} \nonumber \\
 &+& \int_{-(V_0
-\varepsilon/v)}^{(V_0-\varepsilon)/v}\frac{d P_x}{2\pi \hbar}
\sqrt{\Big (\frac{V_0-\epsilon}{v}\Big)^2 -P_x^{2}} \Big \} V;
 \label{ehcurrenttotal}
\end{eqnarray}
%%%%%%%%%%%%%%%%%%%%%%%%%%%%%%%%%%
%
%

Therefore, one sees that  the  current flowing along the p-n
junction $J_{total}^{(eh)}$  is a sum of the standard edge state
of
 currents of separated electrons and holes at the separated sample borders.
As it follows from Eq.(\ref{ehcurrenttotal})
 the value of the
  current  $J_{total}^{(eh)}$ does not
depend on the sign of the applied voltage drop $V$.

In conclusion of the section, the current flowing along the p-n
junction is inevitably the sum of two qualitatively different
types of the currents:

1) the current carried by quasiparticles which are a quantum
superposition of electron and hole states; these states are
delocalized along the lateral junction  and quasiparticles in
those states create current $J_{b}^{(eh)}$ (see
Eqs.(\ref{ehspectrum}), \ref{ehbarriercurrent3}).

2) Currents of electrons and holes in the Landau states which do
not interact with the p-n junction. Such quasiparticles move along
closed semiclassical orbits, only parts of those orbits being
inside the above-mentioned stripe. They create   electron and hole
currents $J_L^{(e)}$ and $J_L^{(h)}$.

%In space, these quasiparticles are localized in the direction
%perpendicular to the junction inside the stripe $- 2 R_H ^{(e)}
%\leq y\leq 2 R_H ^{(h)}$, and they delocalized along the latter.

 As one easy sees
these currents flow in the opposite direction to the current
$J_{b}^{(eh)}$ compensating the latter if $V=0$  (see
Eqs.(\ref{ehbarriercurrent3},\ref{eLandaucurrentElectrons},\ref{hLandaucurrentElectrons})).
This statement is correct in the lowest semiclassical
approximation in which all the three currents have been  obtained.
In quantum  oscillating corrections to the
 smooth  part of the currents considered here, as well as  in the quantum Hall
regime  (in which dynamics and kinetics of quasiparticles are of
the fundamentally quantum character) the above-mentioned
compensation is absent because of the different quantum behavior
of electrons in the Landau states and those delocalized along the
junction.  As the quasiparticles  in such a situation are in the
essentially quantum states it seems doubtful whether arising of
the oscillations is a manifestation of the semiclassical
snake-like trajectories (snake states)\cite{Carmier,Beenakker}.
Note that peculiar conductance oscillations were observed in
samples of high quality\cite{Thiti,Rickhaus}, the latter condition
being one of the necessary conditions for observation of quantum
effects.

\section{Giant oscillations of the conductance of grphene ribbon
with  n-p-n lateral junction under magnetic field.
\label{giantoscillations}}
\subsection{Ballistic transport \label{ballisticsubsection}}
%
%
%%%%%%%%%%%%%%%%%%%%%%%%%%%%%%%%%%%%%%%%%%%%%%%
  \begin{figure}
  %%%%%%%%%%%%%%%%%%%%%%%%%%%%%%%% \centerline{\psfig{figure=zlaser2.eps,width=8cm}}
  \centerline{\includegraphics[width=8.0cm]{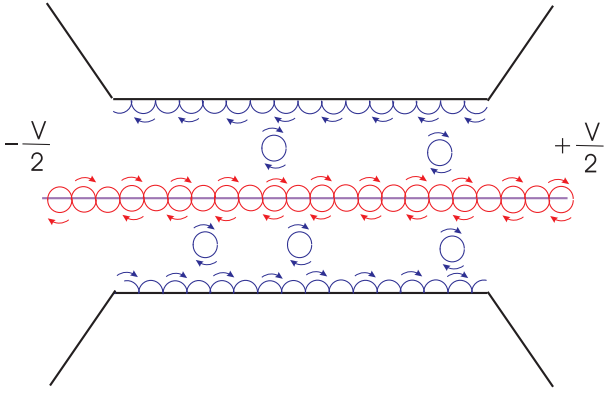}}
  %\vspace{1cm}
  \caption{Schematic presentation of the  graphene ribbon
  with a lateral n-p-n  junction biased by voltage V under magnetic field.}
\label{ehesample}
  \end{figure}
  %%%%%%%%%%%%%%%%%%%%%%%%%%%%%%%%%%%%%%%%%%%%%%%
%
%
In this section the ballistic transport through a graphene ribbon
with an n-p-n junction under magnetic field is considered.

The sample is schematically shown in Fig.\ref{ehesample}. As one
sees there are two qualitatively  different types of currents
flowing along the sample:  current $J^{(ee)}_{edge}$ carried by
electrons  edge states at the external sample boundaries and
current $J_b$ carried by electrons localized along the n-p-n
junction the dispersion equation of which is given by
Eq.(\ref{ehespectrum}) (their spectrum is presented in
Fig.\ref{ee+ehspectra}B)

According to the Landauer-B\"{u}ttiker approach, based on the
relationship between the conductance and the transmission
probability in propagating channels, \cite{Dittrich} the linear
conductance may be written as follows:
%
%
%%%%%%%%%%%%%%%%%%%%%%%%%%%%%%%%%%
\begin{eqnarray}
G&=&\frac{2 e^2}{k T}\sum_n \int_{-\varepsilon/v}^{\varepsilon/ v} \frac{d P_x}{2\pi \hbar}
\Big|v^{(ee)}_x\left[\varepsilon^{(ee)}_n(P_x),P_x\right]\Big|\nonumber \\
&\times& \cosh^{-2}
\frac{\varepsilon^{(ee)}_n(P_x)-\varepsilon_F}{2kT}; \label{G1}
\end{eqnarray}
%%%%%%%%%%%%%%%%%%%%%%%%%%%%%%%%%%
%
%
where the  quasiparticle velocity is $v^{(ee)}_x=d
\varepsilon^{(ee)}_n /d P_x$

In the analogous way as deriving DOS, Eq.(\ref{NuGeneral}), one
gets the conductance along then-p-n lateral junction as follows
(details of calculations are presented in Appendix \ref{Gappend}):

%%%%%%%%%%%%%%%%%%%%%%%%%%%%%%%%%%
\begin{eqnarray}
\frac{G_{b}\left(H\right)}{G_{edge}}=\frac{4}{\pi}\sum_n\int_{-1}^{1}dq\sqrt{1-q^2}\nonumber \\
\Big(\tanh\frac{\varepsilon_n^{(t)}-\varepsilon_F}{2 k T}
-\tanh\frac{\varepsilon_n^{(b)}-\varepsilon_F}{2 k T}\Big)
\label{GcleanH}
\end{eqnarray}
%%%%%%%%%%%%%%%%%%%%%%%%%%%%%%%%%%
%
%
where $G_{edge}= G_0 (2R_H/\lambda_F)$ is  the conductance of the
edge states in the graphene ribbon in the absence of the lateral
junction, $G_0 =e^2/h$ is  the conductance quant and
$2R_H/\lambda_F$is the number of the propagating channels of the
edge states; $\lambda_F=\hbar/p_F$ while
$\varepsilon_n^{(b,t)}(H)$ are the bottom and the top of the
$n$-th electron energy band which are found from the condition
$\cos{\Phi_{+}}(\varepsilon) = |r(P_x)|$ (see
Eq.(\ref{ehespectrum})):
%
%
%%%%%%%%%%%%%%%%%%%%%%%%%%%%%%%%%%
\begin{eqnarray}
\varepsilon_n^{(b,t)}(H)&=& v \sqrt{\frac{e \hbar H}{c}}\sqrt{2
n+1}
\nonumber \\
&\pm& \frac{v^2}{2 \varepsilon_F}\frac{e \hbar
H}{c}\Big(1-\frac{2}{\pi} \arccos{|r(P_x)|} \Big)
\label{GapBoundary}
\end{eqnarray}
%%%%%%%%%%%%%%%%%%%%%%%%%%%%%%%%%%
%
%

\begin{figure}
  %%%%%%%%%%%%%%%%%%%%%%%%%%%%%%%% \centerline{\psfig{figure=zlaser2.eps,width=8cm}}
  \centerline{\includegraphics[width=8.0cm]{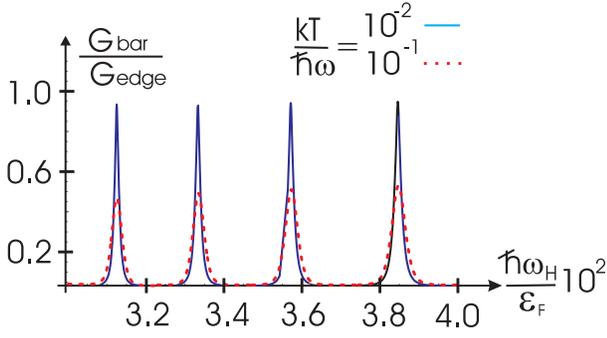}}
  %\vspace{1cm}
  \caption{Conductance oscillations with variations of the magnetic field;
  $G_{edge}=(e^2/h)(2 R_h/\lambda_F)$
is the conductance of  edge states in the graphene ribbon in the
absence of the lateral junction. Numerical calculations are
performed
  for the semiclassical parameter $\eta =10^{-2}$ and the n-p-n reflection probability parameter
  $\lambda =0.2$.}\label{GpureH}
  \end{figure}

The dependance of the conductance along the n-p-n junction on
magnetic filed is shown in Fig.\ref{GpureH}.

%
%
%%%%%%%%%%%%%%%%%%%%%%%%%%%%%%%%%%%%%%%%%%%%%%%
  %%%%%%%%%%%%%%%%%%%%%%%%%%%%%%%%%%%%%%%%%%%%%%%
%
%

\begin{figure}
  %%%%%%%%%%%%%%%%%%%%%%%%%%%%%%%% \centerline{\psfig{figure=zlaser2.eps,width=8cm}}
  \centerline{\includegraphics[width=8.0cm]{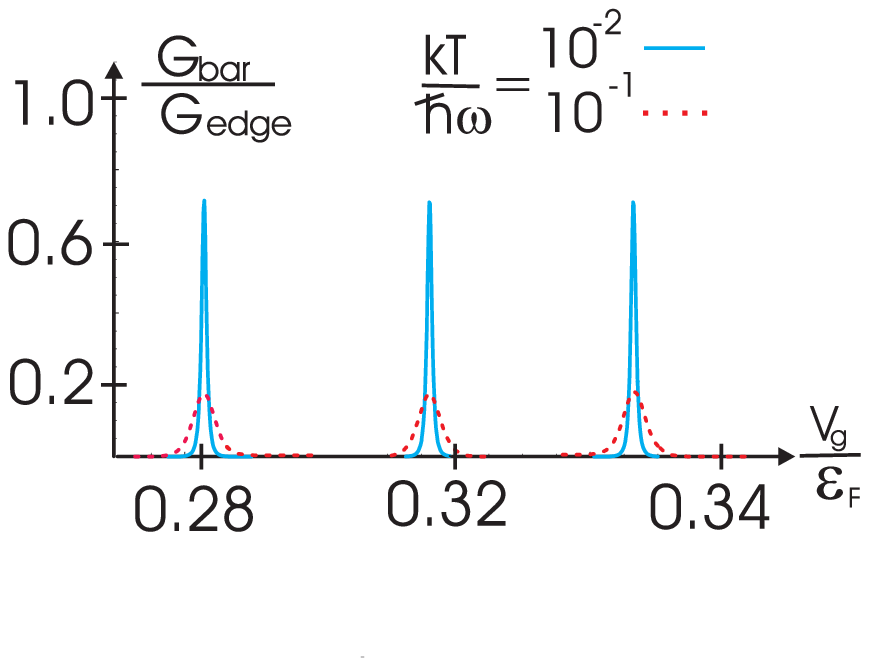}}
  %\vspace{1cm}
  \caption{Conductance oscillations with variations of the gate voltage $V_g$;
  $G_{edge}=(e^2/h)(2 R_h/\lambda_F)$
is the conductance of  edge states in the graphene ribbon in the
absence of the lateral junction. Numerical calculations are
performed
  for the semiclassical parameter $\eta =10^{-2}$ and the n-p-n reflection probability parameter
  $\lambda =0.2$.}
 \label{GpureVg}
  \end{figure}

Giant oscillation of the conductance at a fixed magnetic field $H$
may be observed if  the chemical potential  is varied  together
with the gate potential $V_g$. In this case the conductance is
determined by Eq.(\ref{GcleanH}) in which $\varepsilon_F$ is
changed to $\varepsilon_F + eV_g$. This dependence of the
conductance on the gate potential is presented in
Fig.\ref{GpureVg}.
%
%
%%%%%%%%%%%%%%%%%%%%%%%%%%%%%%%%%%%%%%%%%%%%%%%
  %%%%%%%%%%%%%%%%%%%%%%%%%%%%%%%%%%%%%%%%%%%%%%%
%
%
%

\begin{figure}
  %%%%%%%%%%%%%%%%%%%%%%%%%%%%%%%% \centerline{\psfig{figure=zlaser2.eps,width=8cm}}
  \centerline{\includegraphics[width=8.0cm]{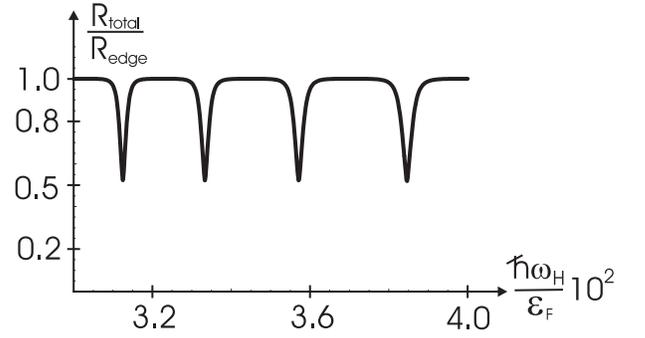}}
  %\vspace{1cm}
  \caption{Total resistance of a graphene ribbon with one n-p-n
  interface under magnetic field. $R_{edge}$ is the resistance of the graphene
 ribbon in the absence of the lateral junction. An extremely small change of the
 magnetic field (or the voltage gate) may controllably cause the 50\% change of
 the total ribbon resistance.Numerical calculations are
performed
  for the semiclassical parameter $\eta =10^{-2}$ and the n-p-n reflection probability parameter
  $\lambda =0.2$.}
\label{ResistanceTotal}
  \end{figure}

The total current flowing along a graphene ribbon with an e-h-e
lateral junction (see Fig.\ref{ehesample}) is $J_{total}=J_{bar}
+2 J_{edge}$ where  $J_{bar}$ is the current carried by electrons
skipping along the junction and $2 J_{edge}$ are the edge state
currents. This current may be written as $J_{total} =V/ R_{total}$
where $R_{total}$ is the total resistance of the ribbon. For a
ribbon  with $N$  parallel lateral junctions its total resistance
is
%
%%%%%%%%%%%%%%%%%%%%%%%%%%%%%%%%%%
\begin{eqnarray}
R_{total} =\frac{1}{ G_{edge}+N G_{b}} \label{RtotalEquation}
\end{eqnarray}
%%%%%%%%%%%%%%%%%%%%%%%%%%%%%%%%%%
%
%
This equation is written under assumption that the distance
between the junctions $L_b > 2 R^{(ee)}_H$ and the width of the
ribbon $W
> 2 R^{(ee)}_H  N  $.
Numerical calculations of the total resistance $R_{total}$ for
$N=1$ with the use of Eq.(\ref{GcleanH}) is presented in
Fig.\ref{ResistanceTotal}.
%
%%%%%%%%%%%%%%%%%%%%%%%%%%%%%%%%%%%%%%%%%%%%%%%
  %%%%%%%%%%%%%%%%%%%%%%%%%%%%%%%%%%%%%%%%%%%%%%%
%
%
As one sees from Eqs.(\ref{GpureH},\ref{RtotalEquation}) and
Fig.\ref{ResistanceTotal} a variation of the magnetic field
$\delta H/H \approx 10^{-1} \hbar \omega_H/\varepsilon_F \ll 1$
produces a 50\% jump of  the total resistance of the ribbon with
one lateral junction. As one readily sees the resistance jump for
the ribbon  with $N$ lateral junctions is
%
%%%%%%%%%%%%%%%%%%%%%%%%%%%%%%%%%%
\begin{eqnarray}
\frac{R_{total}^{(max)}-R_{total}^{(min)}}{R_{total}^{(max)}}=
\frac{N}{(N+1} \label{RtotalN}
\end{eqnarray}
%%%%%%%%%%%%%%%%%%%%%%%%%%%%%%%%%%
%
%
that allows to have the giant magnetoresistance controlled be
small variation of either the magnetic field or the gate voltage
(here $R_{total}^{max,min}$ are the maximal and minimal values of
the total resistance. For $N=3$, e.g., the jump is  75\% of the
total resistance. This property of such  electrically gated
graphene ribbons may be useful in modelling of devices based on
the giant magnetoresistance effects of other types.

In the next subsection the current flowing along a graphene ribbon
with a lateral n-p-n junction under magnetic field and in the
presence of impurities is considered.

\subsection{Dissipative  transport.}

As in the case of the magnetic breakdown phenomenon, dynamic and
kinetic properties of quasi-particles skipping along the junction
under magnetic field are of the fundamentally quantum mechanical
nature due to the quantum interference of their wave functions
with semiclassically large phases. Thus, in order to analyze the
transport properties of the  quasi-particles in the presence of
impurities  it is convenient to start with the the equation for
the density matrix $\hat{\rho}$ in the $\tau$ approximation:
%
%
%%%%%%%%%%%%%%%%%%%%%%%%%%%%%%%%%%
\begin{eqnarray}
\frac{i}{\hbar}\Big[\hat{\rho},\hat{H_0}\Big]-\frac{i}{\hbar}\Big[\hat{\rho},e{\cal
E}\hat{x }\Big]+\frac{\hat{\rho} - f_0(H)}{t_0}=0;
 \label{matrixEq}
\end{eqnarray}
%%%%%%%%%%%%%%%%%%%%%%%%%%%%%%%%%%
%
%
Here, $\hat{H}$ is the Hamiltonian corresponding to the
Schr\;odinger equation Eq.(\ref{Schroedinger}), $f_0$ is the Fermi
distribution function, ${\cal E}$ is the electric field along the
 junction, $t_0 $ is the electron scattering time.

Writing the density matrix in the form $\hat{\rho}= f_0(\hat{H})+
\hat{\rho}^{(1)}$ and linearizing Eq.(\ref{matrixEq}) with respect
to the electric field  one gets
%
%
%%%%%%%%%%%%%%%%%%%%%%%%%%%%%%%%%%
\begin{eqnarray}
\frac{i}{\hbar}\Big[\hat{\rho}^{(1)},\hat{H_0}\Big]+\frac{\hat{\rho}^{(1)}
}{t_0}=- e {\cal E}\hat{v}_x f_0^{\prime}(\hat{H});
 \label{matrixEqLinear}
\end{eqnarray}
%%%%%%%%%%%%%%%%%%%%%%%%%%%%%%%%%%
%
%
where $\hat{v}_x$ is the quantum mechanical operator of the
quasi-particle velocity projection on the electric field
direction, $f_0^{\prime}(\varepsilon) =d f(\varepsilon)/d
\varepsilon$.

In terms of the density matrix the current carried by the
electrons skipping along the junction is written as follows:
%
%
%%%%%%%%%%%%%%%%%%%%%%%%%%%%%%%%%%
\begin{eqnarray}
J=2 e Tr\left\{\hat{v}_x \hat{\rho}\right\}
 \label{currenttrace}
\end{eqnarray}
%%%%%%%%%%%%%%%%%%%%%%%%%%%%%%%%%%
%
%

Taking the matrix elements of  equation Eq.(\ref{matrixEqLinear})
with respect  to proper functions of Schr\"{o}dinger equation
Eq.(\ref{Schroedinger}) written in the Dirac notations
%
%
%%%%%%%%%%%%%%%%%%%%%%%%%%%%%%%%%%
\begin{eqnarray}
\hat{H}\left|\kappa\right \rangle=\varepsilon_\kappa
\left|\kappa\right \rangle;
 \label{DiracNotation}
\end{eqnarray}
%%%%%%%%%%%%%%%%%%%%%%%%%%%%%%%%%%
%
%
 (here $\kappa =\{n,P_x\}$) one finds the density matrix. Inserting the found solution in
Eq.(\ref{currenttrace}) one obtains the current $J$ as follows:
%
%
%
%%%%%%%%%%%%%%%%%%%%%%%%%%%%%%%%%%
\begin{eqnarray}
J=-e^2 {\cal E} \sum_{\kappa, \bar{\kappa}} \frac{\partial
f_0}{\partial \varepsilon}\Big|_{\varepsilon=\varepsilon_\kappa}
\frac{\hbar\big|v_{\kappa,\bar{\kappa}}\big|^2}{i
\big(\varepsilon_\kappa -\varepsilon_{\bar{\kappa}}\big) +\hbar
\nu_0}
 \label{currentmatrix}
\end{eqnarray}
%%%%%%%%%%%%%%%%%%%%%%%%%%%%%%%%%%
%
where $\sum_{\kappa}= \sum_n \int d P_x/2\pi \hbar $ while
$v_{\kappa,\bar{\kappa}}=<\bar{\kappa}|\hat{v}_x|\kappa>$ and
$\nu_0 =1/t_0$ is the electron-impurity relaxation frequency.

As follows from Eq.(\ref{ehespectrum})(see also
Fig.\ref{ee+ehspectra}B) the distance between energy levels
$\big|\varepsilon_\kappa -\varepsilon_{\bar{\kappa}}\big| \sim
\hbar \omega_H$ and hence for the case considered below $ \omega_H
\gg \nu_0$ the main contribution to the sum is of the diagonal
elements because the diagonal element $v_{\kappa,\kappa} \neq 0$
for delocalized quasi-particles. From here it follows that the
current along the junction  may written as

%
%
%%%%%%%%%%%%%%%%%%%%%%%%%%%%%%%%%%
\begin{eqnarray}
J=-e^2 {\cal E}t_0 \sum_{n}
\int_{-\varepsilon_F/v}^{\varepsilon_F/v}\frac{d P_x}{2\pi
\hbar}\big|v^{(ee)}_n (P_x)\big|^2 \frac{\partial f_0}{\partial
\varepsilon}\Big|_{\varepsilon=\varepsilon^{(ee)}_n(P_x)}
 \label{currentmatrix}
\end{eqnarray}
%%%%%%%%%%%%%%%%%%%%%%%%%%%%%%%%%%
where $v^{(ee)}_n(P_x)\equiv v_{\kappa,\kappa} =d
\varepsilon^{(ee)}_n(P_x)/dP_x $.

Using the same approach as in Subsection \ref{ballisticsubsection}
one finds the conductance along the n-p-n junction in the presence
of impurities as follows:
%
%
%%%%%%%%%%%%%%%%%%%%%%%%%%%%%%%%%%
\begin{eqnarray}
\frac{G_{dirty}}{G_{Drude}}=-8 \int d \varepsilon \frac{\partial
f_0(\varepsilon -e V_g)}{\partial \varepsilon}
\Big| \sin{\Phi_{+}(\varepsilon)}\Big| \int_{-1}^{1}d \xi \nonumber \\
 \frac{(1-\xi^2)\sqrt{r^2(\xi)-  \cos^2{\Phi_{+}(\varepsilon)}}\theta\Big[|r^{(ee)}|^2(\xi)-
 \cos^2{\Phi_{+}(\varepsilon)}\Big]}{\pi^2\sin^2{\Phi_{+}(\varepsilon)}-\big(|r^{(ee)}|^2(\xi)-
 \cos^2{\Phi_{+}(\varepsilon)\big)}\big(2 \arcsin{\xi}\big)^2};
\label{eheGdirty}
\end{eqnarray}
%%%%%%%%%%%%%%%%%%%%%%%%%%%%%%%%%%
where $G_{Drude}= \sigma_0 R_H$. Here  $\sigma_0= \varepsilon_F
e^2 t_0/\hbar^2$ is the Drude conductivity of graphene in the
absence of magnetic field, $H=0$.

\begin{figure}
  %%%%%%%%%%%%%%%%%%%%%%%%%%%%%%%% \centerline{\psfig{figure=zlaser2.eps,width=8cm}}
  \centerline{\includegraphics[width=8.0cm]{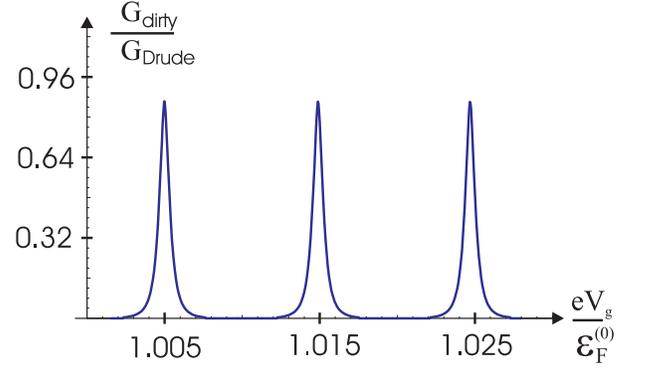}}
  %\vspace{1cm}
  \caption{Giant oscillations of the conductance of a dirty
  graphene ribbon with an n-p-n interface
under  variations of the gate voltage $V_g$ normalized to
$G_{Drude}=\sigma R_H^{(e)}$. Numerical calculations are performed
  for the semiclassical parameter $\eta =10^{-2}$ and the n-p-n reflection probability parameter
  $\lambda =0.2$.}
\label{Gdirty}
  \end{figure}

Dependence of the conductance  on the gate voltage $V_g$ in the
presence of impurities is presented in Fig.\ref{Gdirty}. As one
sees the conductivity $G_{dirty}/R_H$ reaches the Drude
conductivity when the energy $\varepsilon_F +e V_g$ is in the
middle of a band and is equal to zero when it is inside a gap of
the energy spectrum (see Fig.\ref{ee+ehspectra}B).

The above giant oscillations of the conductance
 are based on the
quantum interference of the edge states on the both sides of the
lateral n-p-n junctions that transforms the gapless spectra of the
separated edge states into a series of alternating  energy gaps
and bands.  In the same way as it takes place for magnetic
breakdown this pure quantum mechanical picture holds if the path
traversed by the "new" quasiparticle between collisions is greater
than the individual classical trafectory \cite{Lifshitz}. It means
that in the case under consideration the bands give the main
contribution to the conductance if the following inequality holds:
%
%
%%%%%%%%%%%%%%%%%%%%%%%%%%%%%%%%%%
\begin{eqnarray}
<v_{gr}> t_0 \gg R_H^{(e)}
 \label{quantumcondition}
\end{eqnarray}
%%%%%%%%%%%%%%%%%%%%%%%%%%%%%%%%%%
where  $<...> =\int_0^{p_F}(...)dP_x/p_F$, the group velocity
$v_{gr}=|r^{(ee)}(P_x)|d \varepsilon^{(ee)}_n /d P_x$ and $t_0$ is
the free path time, $|r(P_x)|$ is the probability amplitude of the
reflection at the n-p-n junction. This inequality may be
re-written as
%
%
%%%%%%%%%%%%%%%%%%%%%%%%%%%%%%%%%%
\begin{eqnarray}
<|r^{(ee)}(P_x)|> \gg \frac{R_H}{l_0}
 \label{quantumcondition}
\end{eqnarray}
%%%%%%%%%%%%%%%%%%%%%%%%%%%%%%%%%%
where $l_0=v t_o$ is the free path length.

%%%%%%%%%%%%%%%%%%%%%%%%%%%%%%%%%%%%%%%%%%%%%%%%%%%%%%%%%%%%%%%%%%%%%%%%%%%%%%%%%%%%%%%%%%%%%%
%%%%%%%%%%%%%%%%%%%%%%%%%%%%%%%%%%%%%%%%%%%%%%%%%%%%%%%%%%%%%%%%%%%%%%%%%%%%%%%%%%%%%%%%%%%%%%
%%%%%%%%%%%%%%%%%%%%%%%%%%%%%%%%%%%%%%%%%%%%%%%%%%%%%%%%%%%%%%%%%%%%%%%%%%%%%%%%%%%%%%%%%%%%%%
%%%%%%%%%%%%%%%%%%%%%%%%%%%%%%%%%%%%%%%%%%%%%%%

%
%
  %%%%%%%%%%%%%%%%%%%%%%%%%%%%%%%%%%%%%%%%%%%%%%%
%
%

\section{Discussion and Conclusion.}

Quantum dynamics and kinetics of quasipartcles in a graphene
ribbons with either n-p or n-p-n lateral interface under magnetic
field is considered in the semiclassical approximation.
Calculations of the  current flowing along the n-p junction in the
voltage biased ribbon  show that there are three different
currents inside the regions $-2 R_H^{(e)} \leq y < 0$ and $0< y
\leq 2 R_H^{(h)}$ around the lateral junction at $y=0$ (see
Fig.\ref{p-ncurrents}). One of them is the current  of the
quasiparticles  skipping along the interface, $J_b^{(eh)}$, (see
Eq.(\ref{ehbarriercurrent3})). The other two are currents
$J_L^{(e)}$ and $J_L^{(h)}$ which are  created by quasiparticles
in the localized Landau states the closed orbits of which are
partially inside the above-mentioned regions around the n-p
lateral junction (see
Eqs.(\ref{eLandaucurrentElectrons},\ref{hLandaucurrentElectrons}).
The latter currents flow in the opposite direction to the current
of the skipping quasiparticles $J_b^{(eh)}$ exactly compensating
it in the absence of the bias voltage. As a result, the measurable
current (which is the sum of those three currents) flowing along
the biased n-p junction is the sum of   two  standard edge state
currents of  electrons and holes independently flowing along the
junction (see Eq.(\ref{ehcurrenttotal})). Therefore, the
snake-like states   suggested in Ref.\cite{Beenakker} do not
manifest themselves in the main smooth part of the conductance of
a graphene ribbon with an n-p interface. In principle, the
snake-like states  could
 implicitly affect  the quantum oscillating corrections or the
conductance in the regime of the quantum Hall effect but the
essentially quantum character of the latter contradicts  the
classical nature of the former.

It is also shown that giant conductance oscillations may arise in
a biased graphene ribbon with an n-p-n lateral interface under
magnetic field. In such a state of the ribbon, depending on the
number of n-p-n interfaces inside the ribbon, its total
magnetoresistance may be controllably changed  by $50\%\div 90\%$
by an extremely small variation of the gate voltage or the
magnetic field  (see Fig.\ref{ResistanceTotal} and
Eq.(\ref{RtotalN}))

\appendix
\section{Dispersion equation for quasiparticles skipping along an n-p junction
under magnetic field. \label{matching}}

 The semiclassical solutions of Eq.(\ref{Schroedinger})
above and below  the junction ($0<y<y_h$ and $y_e <y<0$,
respectively) are
%
%
%%%%%%%%%%%%%%%%%%%%%%%%%%%%%%%%%%
\begin{eqnarray}
&\widehat{\Psi}_h&=\frac{C_h}{(y_t-y)^(1/4)} \nonumber \\
&\times & \Big[ \left(\begin{matrix} 1\\
-e^{- i\varphi_h}
\end{matrix} \right) \exp\left\{\frac{i}{\hbar}\int_y^{y_h}p_h(y^\prime)dy^\prime)-\frac{\pi}{4}\right\} +
h.c.\Big], \nonumber \\
&\widehat{\Psi}_e&=\frac{C_e}{(y_t-y)^(1/4)} \nonumber \\
&\times & \Big[ \left(\begin{matrix} 1\\
-e^{- i\varphi_e}
\end{matrix} \right) \exp\left\{\frac{i}{\hbar}\int_{y_e}^y p_e(y^\prime)dy^\prime)-\frac{\pi}{4}\right\} +
h.c.\Big]
 \label{wavefunctions}
\end{eqnarray}
%%%%%%%%%%%%%%%%%%%%%%%%%%%%%%%%%%
%

where
%
%
%%%%%%%%%%%%%%%%%%%%%%%%%%%%%%%%%%
\begin{eqnarray}
y_h&=&\frac{c}{eH}\left(\frac{V_0 -\varepsilon}{v}-P_x\right); \varphi =\arctan\frac{p_h(y)}{P_x+eH y/c}\nonumber \\
p_h(y)&=&\sqrt{\left(\frac{V_0-\varepsilon}{v}\right)^2-\left(P_x+\frac{eH}{c}y\right)^2}\nonumber
\\
y_e&=&-\frac{c}{eH}\left(\frac{\varepsilon}{v}+P_x\right); \varphi_e =\arctan\frac{p_e(y)}{P_x+eHy/c}   \nonumber \\
p_e(y)&=&\sqrt{\left(\frac{\varepsilon}{v}\right)^2-\left(P_x+\frac{eH}{c}y\right)^2}
\label{momentaandphi}
\end{eqnarray}
%%%%%%%%%%%%%%%%%%%%%%%%%%%%%%%%%%
%
%
are the turning points while $P_x$ is the conserving generalized
momentum.

The constants $C_h$ and $C_e$ are determined by the matching of
the above wave functions at the lateral  junction and by the
normalization condition.

In order to match the wave functions at the junction, $y=0$, it is
convenient to  write the integrals in Eq. (\ref{wavefunctions}) as
$\int_y^{y_t,y_b}(...)dy^\prime\approx
\int_0^{y_t,y_b}(...)dy^\prime + \int_y^0...dy^\prime)$. After
expanding  the latter integrals  in $|y|/ R_H \ll 1 $ one gets
%
%
%%%%%%%%%%%%%%%%%%%%%%%%%%%%%%%%%%
\begin{eqnarray}
\int_y^{y_h}p_h(y^\prime)\frac{dy^\prime}{\hbar}\approx \frac{
S_h}{e\hbar H/c} - \frac{y p_h(0)}{\hbar}, \nonumber \\
\int_{y_{e}}^{y}p_e(y^\prime)\frac{dy^\prime}{\hbar}\approx \frac{
S_b}{e\hbar H/c} + \frac{y p_e(0)}{\hbar} \label{phaseexpansion}
\end{eqnarray}
%%%%%%%%%%%%%%%%%%%%%%%%%%%%%%%%%%
%
%
where
%
%
%%%%%%%%%%%%%%%%%%%%%%%%%%%%%%%%%%
\begin{eqnarray}
S_h=
\int_0^{(V_0-\varepsilon/v-P_x)}\sqrt{\left(\frac{V_0-\varepsilon}{v}\right)^2-\left(P_x+
\overline{p}_x\right)^2}d\overline{p}_x, \nonumber \\
S_e=
\int^0_{-(\varepsilon/v+P_x)}\sqrt{\left(\frac{\varepsilon}{v}\right)^2-\left(P_x+
\overline{p}_x\right)^2}d\overline{p}_x
\end{eqnarray}
%%%%%%%%%%%%%%%%%%%%%%%%%%%%%%%%%%
%
%
are the areas of the semiclassical  orbits in the momentum space
shown in Fig.\ref{Areas} in which  $\overline{p}_y
=\sqrt{(\varepsilon/v)^2-\left(P_x+ \overline{p}_x\right)^2}$

As one easily sees from Eq.(\ref{wavefunctions}) and
Eq.(\ref{phaseexpansion}), in the vicinity of the junction $|y|\ll
R_H$ the wave functions in  Eq. (\ref{wavefunctions}) are plane
waves:
%
%%%%%%%%%%%%%%%%%%%%%%%%%%%%%%%%%%
\begin{eqnarray}
&&\widehat{\Psi}_h=\frac{1}{p_h(0)} \nonumber \\
&&\times  \left(\begin{matrix} 1\\
-e^{- i\varphi_h}
\end{matrix} \right) \Big[A_h \exp\left\{i p_h(0)y/\hbar\right\} +B_h \exp\left\{-i p_h(0)y/\hbar\right\}
\Big], \nonumber \\
&&\widehat{\Psi}_e= \frac{1}{p_e(0)} \nonumber \\
&&   \left(\begin{matrix} 1\\
-e^{- i\varphi_e}
\end{matrix} \right) \Big[A_e \exp\left\{i p_e(0)y/\hbar\right\} + B_e \exp\left\{-i p_e(0)y/\hbar\right\}\Big]
 \label{WFexpanded}
\end{eqnarray}
%%%%%%%%%%%%%%%%%%%%%%%%%%%%%%%%%%
%
Here the constants at the plane waves are
%
%
%%%%%%%%%%%%%%%%%%%%%%%%%%%%%%%%%%
\begin{eqnarray}
A_h=  C_h  \exp\{i\big(\frac{
S_h}{e\hbar H/c} +\frac{\varphi_h(0)}{2}-\frac{\pi}{4}\big)\}\nonumber \\
B_h= C_h\exp\{-i\big(\frac{
S_h}{e\hbar H/c} +\frac{\varphi_h(0)}{2}-\frac{\pi}{4}\big)\}\nonumber \\
A_e= C_e \exp\{i\big(\frac{
S_e}{e\hbar H/c} -\frac{\varphi_h(0)}{2}-\frac{\pi}{4}\big)\}\nonumber \\
B_e= C_e \exp\{-i\big(\frac{ S_h}{e\hbar H/c}
-\frac{\varphi_e(0)}{2}-\frac{\pi}{4}\big)\}
\label{plainwavevactors}
\end{eqnarray}
%%%%%%%%%%%%%%%%%%%%%%%%%%%%%%%%%%
%

The incoming quasiparticle undergoes the two-channel scattering at
the n-p junction and hence the constant factors at the scattered
plain waves are
matched with a $2\times 2$ scattering unitary matrix which is written in the general case as%
\begin{equation}
\hat{\tau}^{(eh)} =\left(\begin{matrix} t^{(eh)}& r^{(eh)}\\
-r^{(eh)\ast} & t^{(eh\ast)}
\end{matrix} \right),
\label{taumatrix}
\end{equation}
where $t^{(eh)}$ and $r^{(eh)}$ are the probability amplitudes for
the incoming quasiparticle to pass through and to be scattered
back at the n-p junction, respectively,
$|t^{(eh)}|^2+|r^{(eh)}|^2=1$.

 Using Eqs.\ref{WFexpanded} and  Eq.(\ref{taumatrix})
 one matches the factor
  at the plain waves as follows:
\begin{eqnarray}
B_e=\left(r^{(eh)} A_e +t^{(eh)} A_h\right) \nonumber \\
B_h=\left(-t^{(eh)\ast} A_e +r^{(eh)\ast} A_h\right)
\label{matchequation}
\end{eqnarray}
Replacing $A_{e,h}$ and $B_{e,h}$ by $C_{e,h}$ with the usage of
Eq.(\ref{plainwavevactors}) one finds a $2\times 2$ set of
homogeneous linear algebraic  equations for the required constant
factors $C_{e,h}$ at the semiclassical wave functions
Eq.(\ref{wavefunctions}):
\begin{eqnarray}
\left(e^{-i\theta_e}- r^{(eh)}e^{i\theta_e} \right)C_e
-t^{(eh)}e^{-i\theta_h}C_h=0; \nonumber \\
t^{(eh)\ast}e^{i\theta_e}C_e +\left(e^{i\theta_h}-
r^{(eh)\ast}e^{-i\theta_h} \right)C_h=0;
\label{matchequationfinal}
\end{eqnarray}
where
\begin{eqnarray}
\theta_e=\frac{1}{\hbar}\int_{y_e}^{0}p_e dy; \;\;\;
\theta_e=\frac{1}{\hbar}\int_{y_e}^{0}p_e dy;
 \label{npareas}
\end{eqnarray}
Equating the determinant of equation Eq.{\ref{matchequationfinal}}
to zero one finds the dispersion equation Eq.(\ref{ehspectrum}) of
the main text.

\section{Derivation of the conductance of pure graphene with n-p-n interface. \label{Gappend}}

It  is convenient to re-write Eq.(\ref{G1}) as follows:
%
%
%%%%%%%%%%%%%%%%%%%%%%%%%%%%%%%%%%
\begin{eqnarray}
G&=&\frac{2 e^2}{k T}\int_{-\infty}^{\infty} d\varepsilon
\int_{-\varepsilon/v}^{\varepsilon/ v} \frac{d P_x}{2\pi
\hbar}\cosh^{-2}
\big[\frac{\varepsilon-\varepsilon_F}{2kT}\big]  \nonumber \\
&\times&\Big|v^{(ee)}_x\left(\varepsilon,P_x\right)\Big| \sum_n
\delta\big[\varepsilon-\varepsilon^{(ee)}_n(P_x)\big];
 \label{G11}
\end{eqnarray}
%%%%%%%%%%%%%%%%%%%%%%%%%%%%%%%%%%
%
%

Using Slutskin's approach (see  the derivation of
Eq.(\ref{NuFinal}) and the equation
%
%
%%%%%%%%%%%%%%%%%%%%%%%%%%%%%%%%%%
\begin{eqnarray}
v^{(ee)}_x=-\frac{\partial D^{(ee)}}{\partial
P_x}\Big/\frac{\partial D^{(ee)}}{\partial \varepsilon}
 \label{GapBoundary}
\end{eqnarray}
%%%%%%%%%%%%%%%%%%%%%%%%%%%%%%%%%%
%
%
 one gets
%
%
%%%%%%%%%%%%%%%%%%%%%%%%%%%%%%%%%%
\begin{eqnarray}
G&=&\frac{2 e^2}{k T}\int d\varepsilon \cosh^{-2}
 \big[\frac{\varepsilon-\varepsilon_F}{2kT}\big] \nonumber \\
&\times&\int \frac{d P_x}{2\pi \hbar}\Big|\frac{\partial
D^{(ee)}}{\partial
P_x}\Big|\delta\Big[D^{(ee)}\left(\varepsilon,P_x\right)\Big]
\label{GApp1}
\end{eqnarray}
%%%%%%%%%%%%%%%%%%%%%%%%%%%%%%%%%%
%
%

Inserting the explicit expression for $D^{(ee)}$ (see
Eq.(\ref{ehespectrum})) in the integrand one finds
%
%
%%%%%%%%%%%%%%%%%%%%%%%%%%%%%%%%%%
\begin{eqnarray}
G=\frac{2 e^2}{k T}\int d\varepsilon \cosh^{-2}
\big[\frac{\varepsilon-\varepsilon_F}{2kT}\big] \int \frac{d
P_x}{2\pi \hbar}\nonumber \\
 \Big|r^{(ee)}\Big|\Big|\sin{\Phi_{-}^{(ee)}}\Big|
 \sqrt{\big(\frac{\varepsilon}{v}\big)^2-P_x^2}\nonumber \\
 \delta\Big[\cos{\Phi_{+}^{(ee)}}
 -r^{(ee)}\cos{\Phi_{-}^{(ee)}}\Big] \label{GApp2}
\end{eqnarray}
%%%%%%%%%%%%%%%%%%%%%%%%%%%%%%%%%%
%
%
Expanding the integrand  into the Fourier series  in $\Phi_{-}$
and taking the zero harmonics of it  (which gives the main
contribution in the integral with respect to $P_x$ because other Fourier
harmonics are fast oscillating functions of $P_x$,
see the
derivation of Eq.(\ref{NuFinal})) one gets
%
%
%%%%%%%%%%%%%%%%%%%%%%%%%%%%%%%%%%
\begin{eqnarray}
G\approx \frac{2 e^2 c}{H (\hbar \pi)^2k
T}\int_{-\infty}^{-\infty}
d\varepsilon\int_{-\varepsilon/v}^{\varepsilon/v}dP_x \nonumber \\
\times \frac{\sqrt{(\varepsilon_F/v)^2-P_x^2}}{\cosh^{2}
\big[\varepsilon-\varepsilon_F/2kT\big]
}\Theta\Big[\big|r^{(ee)}(P_x)\big|^2 -
\cos^2\Phi_{+}(\varepsilon)\Big];
 \label{GApp23}
\end{eqnarray}
%%%%%%%%%%%%%%%%%%%%%%%%%%%%%%%%%%
%
%
where $\Theta(x)$ is the unit step function and  $\Phi_{+}^{(ee)}
= (\pi/2(c/e\hbar H))(\varepsilon/v)^2$, see
Eq.(\ref{ehespectrum}). Taking  the integral with respect to
$\varepsilon$ one gets Eq.(\ref{GcleanH}) of the main text.

%%%%%%%%%%%%%%%%%%%%%%%%%%%%%%%%%%%%%%%%%%%%%%%%%%%%%%%%%%%%%%%%%%%%%%%%
%%%%%%%%%%%%%%%%%%%%%%%%%%%%%%%%%%%%%%%%%%%%%%%%%%%%%%%%%%%%%%%%%%%%%%%%

%\date{\today}

%\begin{abstract}
%\end{abstract}

%\date{\today}

%\begin{abstract}
%\end{abstract}

\end{document}